\documentclass[groupaddress]{revtex4}
\usepackage[]{graphicx}
\usepackage{times}

\begin{document} 
\title{Optical properties of \textit{n}- and \textit{p}-type ZnO thin films --- two different approaches to the impurity distribution inhomogeneity} 


\author{Takayuki Makino}
\affiliation{Department of Material Science, University of Hyogo, Ako-gun, Hyogo, Japan; \\}


\email{makino@sci.u-hyogo.ac.jp}
\altaffiliation{Further author information: (Send correspondence to T. M.)\\T.~Makino.:
Part of the present work was performed at the Photodynamics Research Center, The Institute of Physical and Chemical Research (RIKEN), Sendai, Japan.}


\begin{abstract}
We investigated the optical properties of epitaxial \textit{n}- and \textit{p}-type
ZnO films grown on lattice-matched ScAlMgO$_4$ substrates. Chemical doping yielded a severe inhomogeneity in a statistical distribution of involved charged impurities. Two approaches are adopted to treat the inhomogeneity effects; Monte Carlo simulation technique for the \textit{n}-doped films, and the fluctuation theory for the \textit{p}-ZnO.
The broadening of PL band of \textit{n}-ZnO was significantly larger than predicted by theoretical results in which the linewidths of each individual emissions have been determined mainly from the concentration fluctuation of donor-type dopants by the simulation. Moreover, the rather asymmetrical line shape was observed. To explain these features, a vibronic model was developed accounting for contributions from a series of phonon replicas.
In case of \textit{p}-type ZnO:N, analysis of excitation-intensity dependence of the peak shift of donor-acceptor luminescence with a fluctuation model has also proven the importance of the inhomogeneity effect of charged impurity distribution, as in the case of ZnO:Ga. We extracted the inhomogeneity in the sample and acceptor activation energy prepared under the various growth conditions. It is shown that the theoretical results are in good agreement with the experimental 5-K time-resolved luminescence for the systems in a fluctuation field. Finally, localized-state distributions have been studied in N-doped ZnO thin films by means of transient photocurrent measurement.
\end{abstract}

  \maketitle 

\section{INTRODUCTION}
\label{sect:intro}  

Among widegap II-VI semiconductors, ZnO seems to be one of the most promising
materials for optoelectronic applications. This is due to
its stable excitons, having a large binding energy of 59~meV~\cite{makino29,ohtomo_sst}. It is also desired to
more deeply understand the underlying doping physics~\cite{makino_trans,makino-jjap-2006-1} for the achievement of
a compound-semiconductor-quality \textit{p-n} junction~\cite{nature_mat_tsukazaki}. In our previous work, we reported the observation of intense near-band-edge photoluminescence (PL) for ZnO:Ga at room temperature (RT)~\cite{makino_int_Ga}. It can be easily imagined that a high doping level tends to yield a severe inhomogeneity in a statistical distribution of involved charged impurities. In this work, such an effect is characterized with the Monte Carlo simulation technique~\cite{makino-jap-2006}. The phonon interactions are also discussed with a `vibronic' model~\cite{makino_jap_2005_2}.

Donor-acceptor pair (DAP) photoluminescence (PL) bands have been widely used to characterize acceptor-doped semiconductors. Especially, in wide-gap semiconductors, it has been difficult to obtain \textit{p}-type doped materials. However, the time-resolved PL has not hardly been studied~\cite{zeuner2,xiong1} for ZnO:N. For the case of ZnO, a large amount of nitrogen well in excess of $10^{19}$~cm$^{-3}$ has to be doped to achieve \textit{p}-type conductivity even using one of the best growth techniques that have ever reported [a repeated-temperature modulation (RTM) technique]~\cite{nature_mat_tsukazaki}. It has become apparent that the ``standard'' theory formulated by Thomas, Hopfield, and Augustyniak~\cite{tha1} (hereafter, it is called a ``standard'' THA model) is totally incapable of explaining results of the DAP PL for heavily doped materials probably because this does not take the effects of inhomogeneity into account. In this work, to implement the effect of the fluctuation field inherent to the doped samples, we successfully grew \textit{p}-type-dopants-incorporated films, which are suitable for the experimental verification of the relevant ``fluctuation'' theory.

\section{Experimental Procedures}
\label{sect:expt}
We have grown gallium- and nitrogen-doped ZnO films grown by
laser-molecular-beam epitaxy on lattice-matched ScAlMgO$_4$ substrates.
The detailed methods of fabrication have been given elsewhere~\cite{nature_mat_tsukazaki}. A very complicated growth sequence has been adopted for
the nitrogen doping. We measured the photoluminescence (PL) of the doped ZnO layers in
the band-edge region. The experimental procedures~\cite{makino_int_Ga} were identical to those adopted in our previous study and their details were reported elsewhere~\cite{makino19}. Pulsed excitation was provided for the time-resolved
experiments with the frequency doubled beam
of a picosecond mode-locked Ti:sapphire laser.

\section{Results and Discussion}
\label{sect:res}
\subsection{Optical properties of \textit{n}-type doped ZnO}
\begin{figure}[hbtp]
\includegraphics[width=.95\textwidth]{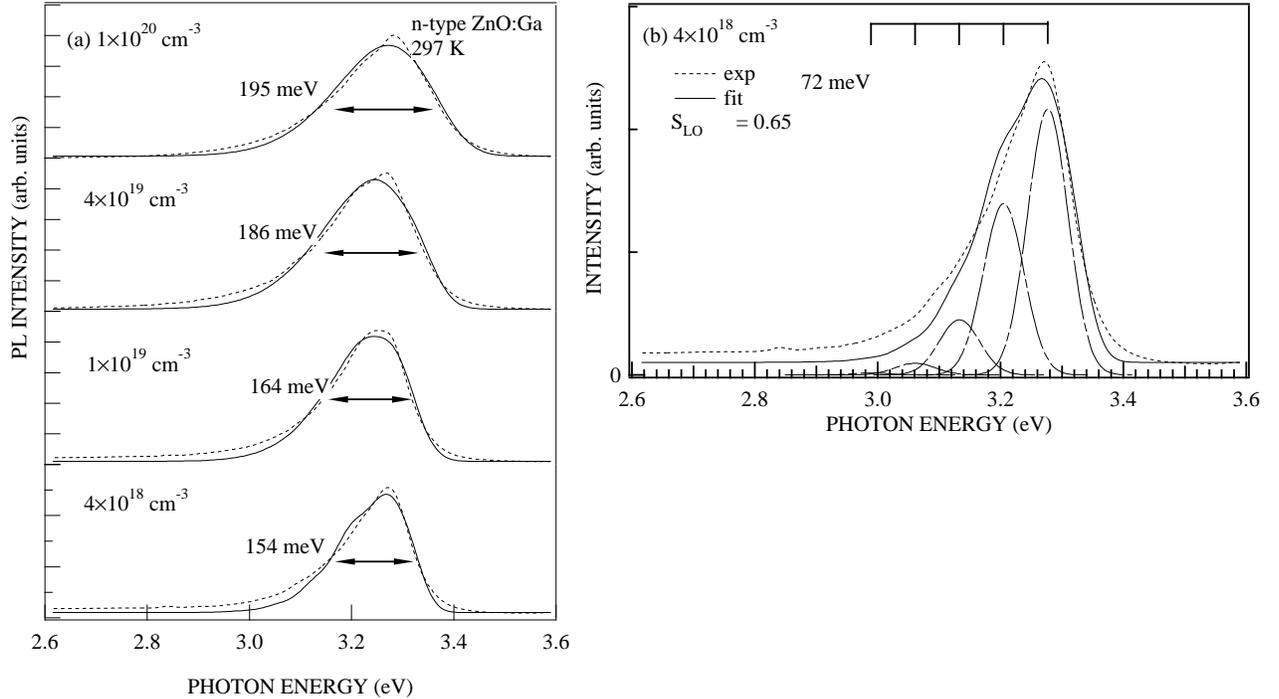}
\caption{(a) Room-temperature photoluminescence spectra
(dashed curves) of \textit{n}-type ZnO doped with different Ga concentrations.
Also shown are the results of the fit to the data using a `vibronic' model (solid lines). (b) The lowermost PL curve of frame (a) with individual contribution
of the emission lines (dash-dotted lines).}
\label{fig:n-type}
\end{figure}

Figure~\ref{fig:n-type}(a) shows PL spectra of ZnO:Ga at four different doping levels (dashed lines). The shape of the PL band is asymmetrical. The linewidths of the PL increase from 154 to 195~meV with an increase in Ga concentration ($n_{Ga}$). The ``theoretical'' broadening for the PL was evaluated in terms of potential fluctuations caused by the random distribution of donor impurities (cf. Fig.~3 of our previous paper~\cite{makino_int_Ga}). These values are listed in Table~I. However, the experimentally observed broadening are greater than the calculated widths ($\sigma_1$).
\begin{table}[hbtp]
\caption{Sample specification and main characteristics for the four \textit{n}-ZnO samples.
Ga concentration ($n_{Ga}$), theoretical broadening $\sigma_1$, broadening
evaluated from the Monte Carlo simulation $\sigma_2$, zero-phonon peak energy $E$, and
the deduced Huang-Rhys factor $S$.}
\begin{center}
		\begin{tabular}{ccccc}
\hline \hline 
$n_{\rm Ga}$ & $\sigma_1$ & $\sigma_2$ & $E$ & $S$\\
(cm$^{-3}$)&(meV)&(meV)& (eV) & \\
\hline

8.0$\times10^{18}$ & 77.7 & 73.5 & 3.277 & 0.64\\
2.0$\times10^{19}$ & 97.1 & 115 &3.272& 0.78\\
8.0$\times10^{19}$ & 150 & 139 &3.297& 1.1\\
1.5$\times10^{20}$ & 188 & 170 &3.345& 1.4\\
\hline \hline 
		\end{tabular}
\end{center}
	\end{table}

\begin{figure}
\begin{center}
\includegraphics[width=.55\textwidth]{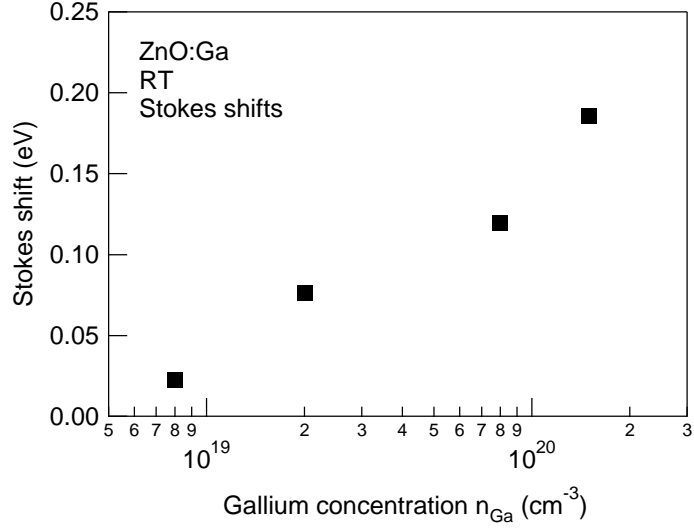}
\end{center}
	\caption{The PL Stokes shifts (closed squares) plotted against the Ga concentration. The measurements were performed at RT.}
	\label{Stokes}
\end{figure}

It is thought that the theoretical widths are exceedingly underestimated because the model might be too
much simplified. We here adopt another approach properly taking the microscopic fluctuation of donor concentration into account; the thermalization redistribution model developed by Zimmermann \textit{et~al}~\cite{zimmermann1}. According to this model, there is a relationship between the Stokes shift and the full-width at half maximum (FWHM) of PL. One can estimate the energy scale of the band potential profile fluctuation from the Stokes shift and predict consistently the FWHM. Figure~\ref{Stokes} shows a Stokes shift of the luminescence plotted against the gallium concentration. Obviously, the shift energy increases with incorporation of the donor dopants. The energy scale of the band potential profile fluctuation can be evaluated from the Stokes shift and predict consistently the FWHM. A Stokes shift of the luminescence is plotted against the gallium concentration as shown in Fig.~\ref{Stokes}.
At sufficiently high temperatures, the above-mentioned relationship is indicated as $E_{Stokes}=-\sigma^2/kT$. The nomenclatures ($k_B$ and $T$) take their conventional meanings. For example, the Stokes shift for the sample with the $n_{Ga}$ of 8.0 $\times 10^{18}$~cm$^{-3}$ is 22~meV, leading to $\sigma $ of 24~meV. It also yields the temperature independent value of 2$\sigma \sqrt(\ln4) \simeq 46$~meV in the FWHM, which can be also supported by the Monte Carlo simulation.

Miller-Abraham's rate for phonon-assisted carrier tunneling between the initial and final states $i$ and $j$ with the energies of $E_i$ and $E_j$ was adopted to simulate the hopping events of photo-created carriers:
\begin{equation}
	\nu_{ij}=\nu_0 \exp \left(- \frac{2r_{ij}}{\alpha} - \frac{(\varepsilon_i-\varepsilon_j+|\varepsilon_i-\varepsilon_j|)}{2kT} \right).
\end{equation}
Here $r_{ij}$ is the distance between the localized states, $\alpha$ is the decay length of the wave function, and $\nu_0$ is the attempt-escape frequency. Hopping was simulated over a randomly generated set of localized states with the sheet density of $N$. Dispersion of the localization energies was in accordance with a Gaussian distribution,
\begin{equation}
	g(\varepsilon ) \propto \exp(- (\varepsilon-E_0)^2/2\sigma^2).
\end{equation}
With the peak positioned at the mean resonant energy $E_0$ and the dispersion
parameter (the energy scale of the band potential profile fluctuation) $\sigma$. The radiative recombination event with the probability $\tau_0^{-1}$ terminates the hopping process. The emission energy at the time of recombination is scored to the luminescence spectrum.
By implementing the thermal broadening effect, the overall FWHM of the PL band is given by:
\begin{equation}
\textrm{FWHM (}\sigma_2\textrm{)} \approx [(2\sigma \sqrt(\ln4))^2+(1.8 k_B T)^2]^{1/2}.
\end{equation}
\begin{figure}[hbtp]
\begin{center}
\includegraphics[width=.7\textwidth]{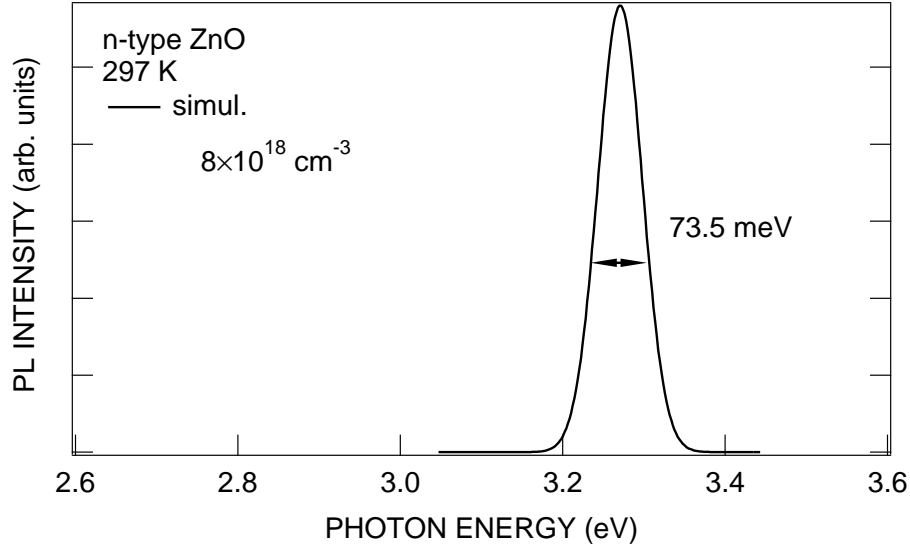}
\end{center}
	\caption{Simulated PL spectrum under stationary excitation. The Ga concentration is shown in the figure.}
	\label{simulated}
\end{figure}
Figure~\ref{simulated} represents, with a solid line, the simulated result obtained for the following values of parameters: $N \alpha^3 =0.1$, $\tau_0 \nu_0=10^4$, and $\sigma \simeq 24$~meV. The lineshape of resulting spectrum was independent of the choices of the $N \alpha^3$ and $\tau_0 \nu_0$ parameters. The simulated curve could be well approximated by the Gaussian function, which is not surprising if considering the measurement temperature (300~K). The FWHMs ($\sigma_2$) of the simulated PL spectra for the samples at four different doping levels are complied in Table~I.

The simulated broadening ($\sigma_2$) is smaller than that of experiments as in the case of the simpler theoretical consideration. This difference is probably due to the contribution of the phonon replicas. In addition, the simulated line-shape is rather symmetrical, which is in great contrast to the asymmetrical shape observed in the experimental spectra. We think that the zero-phonon peak and its phonon replicas are not spectrally resolved well, leading to the larger experimental broadening.
It is inferred that the coupling with longitudinal-optical (LO) phonon is relatively strong. ZnO may have coupling constant significantly stronger than that of GaN or of ZnSe~\cite{makino8}.

Only the terms of longitudinal optical (LO) phonons are taken into account~\cite{reynolds-green}, whose interaction is known to be the strongest. In such a case, the transition energy of the replicas will be $\hbar \omega = E_{\rm ZPL} -  \eta_{\rm LO} E_{\rm LO}$, where $E_{\rm LO}$=72~meV in ZnO, $E_{\rm ZPL}$ is the energy of the zero-phonon peak and $\eta_{\rm LO} $ is an integer.
A Gaussian function with width $\sigma_1$ (cf. Table~I) modeled each individual emission line in the spectrum. The probability of a given phonon emission is proportional to $(S_{\rm LO}^{\eta {\rm LO}}/ \eta_{\rm LO}!)$, where $S_{\rm LO}$ is the Huang-Rhys factor for the LO phonons and $\eta_{\rm LO}$ the number of phonons emitted in a transition. The solid lines in Figs.~\ref{fig:n-type}(a) and (b) were obtained by summing all of the lines using an appropriate set of parameters. Table~I summarizes the donor-impurity concentrations ($n_{\rm Ga}$), zero-phonon peak energies ($E$), and \textit{S}-factors. The respective contributions whose peak positions are shown by five dash-dotted curves in Fig.~\ref{fig:n-type}(b), with equidistant peaks by the $E_{\rm LO}$ of 72~meV. We can say from this fit that the asymmetrical and broad PL band can be explained by the contribution of LO phonon replicas. It should be noted that, even in the undoped ZnO, the zero-phonon emission becomes weaker than LO-phonon-assisted annihilation processes at elevated sample temperatures.
\subsection{Optical properties of \textit{p}-type doped ZnO}
\begin{figure}[hbtp]
\includegraphics[width=.95\textwidth]{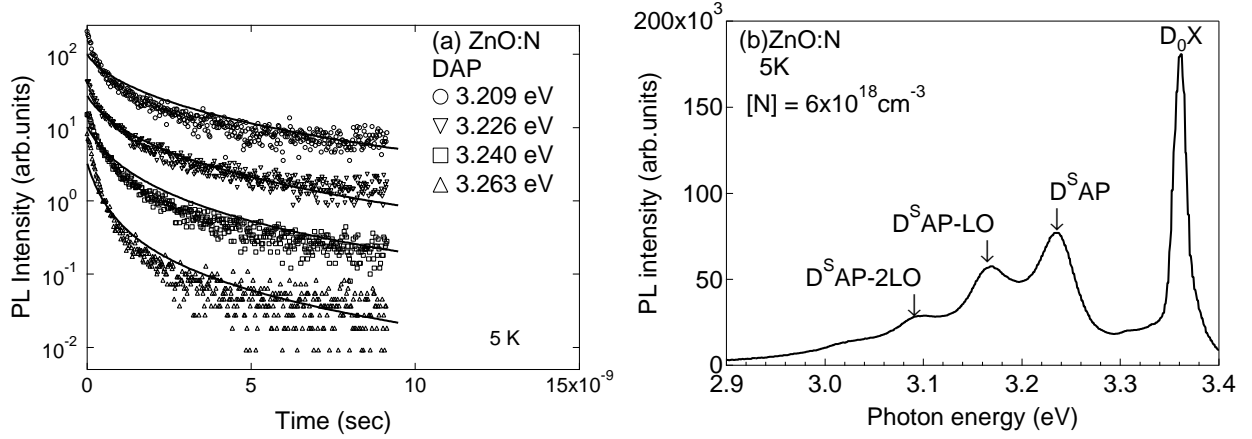}
	\caption{(a): Comparison of the experimental time-resolved PL on ZnO:N at three different emission energies with a fluctuation theory. Open circles are the experimental data, while the solid curves are the theoretical fits with parameters of $\xi = 13$, $\eta = 0.036$, $\tilde{n} = 0.011$, and $W_{\rm max} = 3 \times 10^{10}$~s$^{-1}$. The PL intensities were normalized and each curve was vertically shifted for clarity. The measurement temperature is 5~K. Secondary-ion mass spectroscopy (SIMS) was used for the determinations of the total N concentrations ([N] $\approx 6 \times 10^{18}$~cm$^{-3}$). Its conductivity is not yet \textit{p}-type. Also shown in (b) is the corresponding time-integrated PL.}
	\label{fig:p-type}
\end{figure}
Thin ZnO films are known to be grown more or less under
residual \textit{n}-type. The trial of the \textit{p}-type doping often
results in an observation of the radiative recombination from donor-acceptor
pairs (DAPs). Monte-Carlo simulation~\cite{makino-jap-2006} cannot be used for the analysis of such a DAP system because the photo-carriers in these structures cannot be regarded as a single particle. The experimental PL features cannot be explained with the simple THA model in the case of our N-doped ZnO samples.
Symbols in Fig.~\ref{fig:p-type}(a) show a set of 5-K PL decay curves taken at three different emission energies, while the corresponding time-integrated PL is given in Fig.~\ref{fig:p-type}(b). We first remark that the decay time constants for all the energies are similar with respect to each other and the decay curves are non-exponential. This similarity is difficult to be predicted in the framework of THA theory. In this model, using the pair distance ($R$) between donors and acceptors, the emission energy ($h \nu$) is given by~\cite{tha1}:
\begin{equation}
h \nu = h \nu_\infty+ e^2/\epsilon R = \left( E_G - E_A -E_D \right) + e^2/\epsilon R,
\label{eq:THA}
\end{equation}
where the other symbols take their conventional meanings. Equation~(\ref{eq:THA}) gives a unique relation between $h \nu$ and $R$. The radiative recombination kinetics of DAPs is predominantly determined by the pair distance, $R$. It is well-known that the decay time is strongly dependent on $R$, and hence on $h \nu$. This strong ``dispersion'' effect is in apparent contradiction with our observed data.

According to the treatment of Kuskovsky and his coworkers~\cite{kuskovsky2} in which the fluctuation effects are accounted for, an equation describing the time-dependent PL intensity is now quite different from that in its counterpart~\cite{tha1}.
By averaging over all possible realization of the fluctuation potential,
transient intensity $I_E (t)$ at an energy $E $ is:

\begin{eqnarray}
I_E(\tilde{t}) \propto \frac{W_{\rm max} \tilde{n} \xi^3}{4} 
\exp \left[-\frac{\tilde{n}}{6} {\tilde{t}} \int_0^\infty dx \quad
x^3 \exp (-x-\tilde{t} \exp (-x)) \right] \int_0^\infty 
\frac{du \quad u^{5/2}}{\sqrt{u-1+\exp (-u)}} \cr \times \exp 
\left[ (-\xi u)-\tilde{t} \exp [-\xi u]-\eta  
\frac{(\tilde{E} u-1)^2}{u(u-1+\exp [-u])} \right].
\label{eq:fluct}
\end{eqnarray}
The zero energy as $h \nu_\infty$, defining an effective emission
energy corresponds to $E = h \nu - h\nu_\infty$. We have introduced a
dimensionless time $\tilde{t} = W_{\rm max} t$ and energy $\tilde{E}={\varepsilon Rs E}\slash {e^2}$, as well as parameters $\xi =2 R_s/R_B$, $\eta = e^2 /(4 k T_g \varepsilon R_s)$, and $\tilde{n} = \pi (N_A-N_D) R_B^3$, which have allowed us to have all integrals as dimensionless factor. Here, $N_A$ and $N_D$ are the acceptor and donor concentrations, $R_s$ is the screen radius, and $T_g$ is the frozen temperature. With the use of Eq.~(\ref{eq:fluct}), it is possible to show that the decay is non-exponential, relatively close to stretched-exponential, or even slower,
particularly at very low temperatures.

The solid curves of Fig.~\ref{fig:p-type}(a) refer to the results of theoretical fit at different $h\nu$ according to Eq.~(\ref{eq:fluct}). For the calculation, all the parameters except $\tilde{E}$ were kept fixed ($\xi = 13$, $\eta = 0.036$, $\tilde{n} = 0.011$, and $W_{\rm max} = 3 \times 10^{10}$~s$^{-1}$). These values are quite reasonable, assuming $R_B=23$~$\textrm{\AA}$ (which corresponds to the donor depth of 30~meV) and $R_s=150$~$\textrm{\AA}$. As seen from the figures, the theory results in reasonably satisfactory fits for all $h \nu$'s. The fitting procedure is sensitive to the values of the net acceptor concentration ($N_A-N_D$) and of $W_{\rm max}$.

Although the experimental verification for this theory has been performed in N-doped ZnSe, it should be emphasized that only samples influenced by the very strong fluctuation fields have been analyzed~\cite{makino-jpsj-2006,makino-jpsj-2006-seq}. In regard with this theory, wide applicability on the magnitude of fluctuations has not been yet verified, which is the main focus of the present work. This theory introduces a concept of screening radius ($R_s$) which is intimately related to the degree of fluctuations (smaller $R_s$ corresponds to the severer fluctuations).

We discuss the time-integrated PL [cf. Fig.~\ref{fig:p-type}(b)] from the viewpoint of excitation-dependent shift to quantify $R_s$ of our sample. The PL is dominated by the D$^{\rm S}$AP recombination (D$^{\rm S}$ means a shallower donor) at 3.235~eV. The D$^{\rm S}$AP band exhibits no detectable shift with a change in excitation, suggesting very large radius: $R_s > 100$~$\textrm{\AA}$. We confirmed that the use of $R_s =150$~$\textrm{\AA}$ gives the well-reproduced excitation-intensity dependence of $h \nu$. Here, we used a value of 3.228~eV~\cite{LBzincoxide} as $h \nu_\infty =E_G-E_A-E_D$. $E_G = 3.428$~eV, $E_A = 170$~meV~\cite{lookp-type}, and $E_D = 30$~meV~\cite{look_sst}. On the other hand, the similar analysis for ZnSe:N yielded in: $R_s \sim 20$~$\textrm{\AA}$, indicating that our sample is appropriate to verify whether this theory can also be used to analyze the system with significantly larger $R_s$.

One of the reasons for the unexpectedly short decay compared to those in other semiconductors~\cite{zeuner2} is the difference in $R_s$, because the decay becomes slower with increasing fluctuation. The other one is presumably related to the $W_{\rm max}$, which is one order of magnitude larger than that for ZnSe, and may be inherent to ZnO. Xiong and coworkers~\cite{xiong1} have also reported the DAP PL lifetime of a ZnO:N bulk crystal shorter than many other semiconductors.  The probability $W_{\rm max}$ is represented by the following equation:
\begin{equation}
W_{\rm max} = \frac{128}{3} \frac{e^2 \mu h \nu_\infty E_p}{m_0 \hbar^2 c^3}\times \left[\frac{a_A}{a_D}\right]^3,
\end{equation}
where $\mu $ is the index of refraction, $m_0$ is the electron mass, $c$ is the velocity of light, $E_p $ is a band parameter that has been calculated by Lawaetz using a \textbf{k}$\cdot$\textbf{p} analysis, and the result for ZnO is 28~eV, and the other
symbols take their conventional meanings. The calculation using the effective mass theory for effective Bohr radii of donors and acceptors evaluated the theoretical $W_{\rm max}$ to be: $W_{\rm max}^{th} =4 \times 10^9$~s$^{-1}$, which is one order of magnitude smaller than the experimental value. It should be, however, noted that this theoretical evaluation sensitively depends on the choice of ZnO material parameters adopted in the calculation of the Bohr radii of donors and acceptors. We would like to say that the experimental value of $E_p$ is likely to be greater than 28~eV, the contention of which is based on the comparison of the values with other semiconductors and the strong excitonic effects of ZnO. Note, $E_p$ is related to the matrix element between the conduction and the valence bands.
\subsection{Photoconductivity spectroscopy of N-doped ZnO}
\begin{figure}[hbtp]
\includegraphics[width=.95\textwidth]{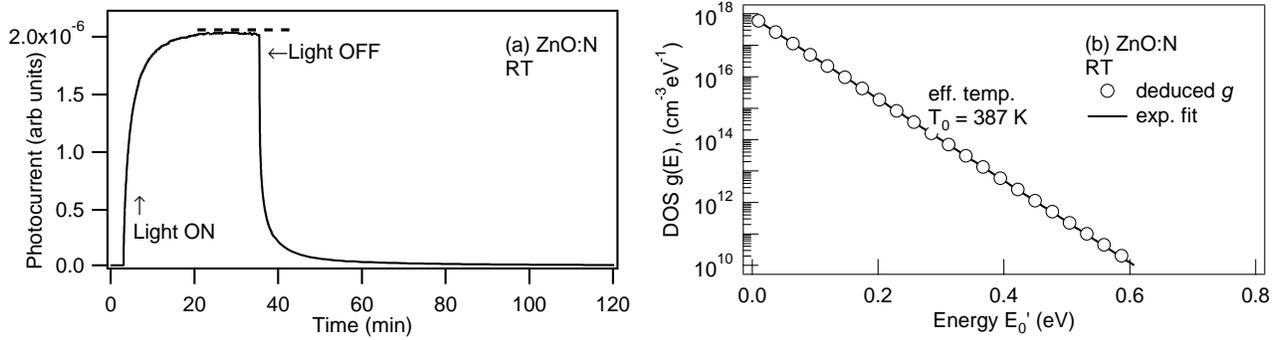}
	\caption{(a) Rise and decay transients of the current in ZnO:N at 293~K. The dashed line stands for a saturated current. (b) Density of states obtained from the PPC transient using
	a high-resolution LTM (open circles) in which the dark current was subtracted. The continuous line corresponds to the result of fit to these circles. }
	\label{fig:photocon}
\end{figure}

The ZnO:N films show a complicated transient behavior of the photoconductance in reaching their respective dark or illuminated equilibrium conduction state. Figure~\ref{fig:photocon}(a) shows a typical current buildup and decay illuminated with sub-band-gap light of 384~nm for $\sim$20 minutes. The photoexcitation as well as the decay of photoexcited carriers are very slow.

After turning off the light, the photo-induced conductivity is observed to prevail for very long times. It is evident that the decay is not essentially exponential for ZnO:N; a rather common feature for persistent photo-current (PPC) experiments, implying interaction between the photo-carriers and localized states during the decay processes. If the PPC is coming from the \textit{AX}-center-like defects, the rising behavior at the edge in PPC excitation spectrum must take the form of $(\hbar \omega - \hbar \omega_{\rm th})^{1.5}$ where $\hbar \omega_{\rm th}$ denotes the threshold energy.

To support this, we calculate the spectral distribution of the trap density-of-states using the multiple trapping model. Resulting rate-equations are given as follows;
\begin{equation}
\frac{dn(t)}{dt} = -\sum_i^m \frac{dn_i(t)}{dt} - \frac{n(t)}{t} + n_0
\delta(t),
\end{equation}
\begin{equation}
\frac{dn_i(t)}{dt} = \omega_i n(t) - \gamma_i n_i(t),
\end{equation}
where $n$ is free carrier density, $n_i$ is carrier density captured at the i-th localized state, $\tau$ is radiative recombination time for free carriers, $n_0$ is photoexcited carrier density, $\omega_i (\propto g(E_i)\times \Delta E$) is a capture rate of carrier, $\gamma_i$ is a releasing rate, $E_i$ is an energy of i-th localized state, and $g$ is density of state.

By solving the Laplace-transformed rate equations analytically, one obtains relationship of $g(E_0^\prime )$ with the Laplace transform of the PPC transient $\hat{I}(t,s)$:
\begin{equation}
g(E_0^\prime ) = \frac{I(0)}{\sigma v k_{\rm B} T} \frac{\hat{I}(s)\hat{K}(s)-2[\hat{J}(s)]^2}{[\hat{I}(s)]^3} s^2,
\end{equation}
\begin{equation}
E_0^\prime  = k_{\rm B} T \ln (\frac{2 \nu }{s}),
\end{equation}
where $\sigma$ is a capturing cross section, $v$ is thermal velocity, $s$ is a Laplace variable, $\nu $ is a detuning frequency, $J(t) = t I(t)$, $K(t) = t^2 I(t)$, and $E_0^\prime $ is the trap energy below the mobility edge. The other symbols used in the equations have the usual meaning.

The LTM thus gives a view of the density-of-state distribution $g$, as was shown in Fig.~\ref{fig:photocon}(b). The spectral lineshape follows an exponentially decaying function. The former component was fitted using an exponential function of the $\exp (-\hbar \omega/ k_{\rm B} T_0)$ type, where $T_0$ means the characteristic temperature. The other symbols have the usual meanings. The best fit to our data is obtained for $T_{0} = 387$~K. Comparing with the results on a polycrystalline ZnO films~\cite{studenikin1}, deduced spectral distribution of $g$ in our case is completely different from that of the polycrystal. In the latter case, $g$ had a peak-like distribution with a maximum near the one-third of the energy gap. This can be attributed to the large difference in the density of grain boundaries~\cite{ohtomo-jjap-2006-1}. The characteristic temperature $T_0$ deduced from Fig.~\ref{fig:photocon}(b) is 387~K.

\section{SUMMARY} 

Gallium and nitrogen were successfully applied as sources of donor and acceptor impurities in the laser molecular-beam epitaxy growth of \textit{n}-type ZnO, respectively. Their optical properties have been investigated in terms of
their effect of inhomogeneous dopant distribution and their phonon interaction.
We also discussed the PL shifting behavior in ZnO:N and compared them with the controls grown by the conventional method. We have explained how the degree of dopant distribution inhomogeneity is extracted from the results of optical
characterization. The activation energy of acceptors has
been evaluated by proper modeling to be $\simeq $170 meV, which
is independent of nitrogen concentration. We have observed transient behavior of photoexcited carriers in semi-insulating N-doped ZnO using photocurrent spectroscopy. The obtained persistent photocurrent shows decay times prevailing over hours. We obtain the spectral distribution of density-of-state distribution of the localized states. The lineshape of the spectrum suggests an existence of localized exciton states. The exponential fit [$\exp (-\hbar \omega/k_{\rm B}T_0)$] deduced a characteristic temperature $T_0$ of 387~K.





\acknowledgments     
 
The authors are strongly thankful to C. H. Chia and Y. Segawa for the fruitful discussions and for providing the experimental data analyzed in this work. Thanks are also due to S. Yoshida, A. Tsukazaki, K. Tamura, A. Ohtomo, M. Kawasaki, and H. Koinuma for providing us the samples and for the encourage throughout the present work. This work was partially supported by MEXT Grant-In-Aid For Scientific Research 18760017, the Asahi Glass Foundation, Iketani Science and Technology Foundation (under contract No. 0181027-A), and the Murata Science Foundation (under contract No. A64104), Japan.



\end{document}